\documentclass{article}

\PassOptionsToPackage{numbers,compress}{natbib}

\usepackage[preprint]{neurips_2026}

\usepackage[utf8]{inputenc} 
\usepackage[T1]{fontenc}    
\usepackage{graphicx}      
\usepackage{hyperref}       
\usepackage{url}            
\usepackage{booktabs}       
\usepackage{makecell}       
\usepackage{enumitem}       
\usepackage{tgpagella}     
\usepackage{amsfonts}       
\usepackage{amsmath}
\usepackage{amssymb}
\usepackage{nicefrac}       
\usepackage{microtype}      
\usepackage{xcolor}         
\usepackage{tcolorbox}      
\tcbuselibrary{skins,listings,breakable}
\usepackage{listings}       
\usepackage{titlesec}       
\titleformat{\section}{\Large\bfseries}{\thesection}{0.8em}{}      
\titleformat{\subsection}{\large\bfseries}{\thesubsection}{0.8em}{}  
\titlespacing*{\section}{0pt}{14pt plus 3pt minus 2pt}{7pt plus 2pt}
\titlespacing*{\subsection}{0pt}{12pt plus 3pt minus 2pt}{5pt plus 2pt}

\usepackage{caption}        
\DeclareCaptionFont{eightpt}{\fontsize{8pt}{10pt}\selectfont}
\hypersetup{colorlinks=true,linkcolor=blue!50!black,citecolor=blue!50!black,urlcolor=blue!50!black}

\newtcblisting{databox}{%
  enhanced, breakable,
  colback=black!3!white, colframe=black!45!white,
  boxrule=0.5pt, arc=1.5pt,
  left=8pt, right=8pt, top=5pt, bottom=5pt,
  listing only,
  listing options={basicstyle=\small\ttfamily, columns=fullflexible,
                   keepspaces=true, breaklines=true}
}

\newtcolorbox{insightbox}[1]{%
  enhanced, breakable,
  colback=blue!8!white, colframe=blue!80!black!80,
  boxrule=0.5pt, arc=1.5pt,
  left=8pt, right=8pt, top=5pt, bottom=6pt,
  title={#1}, fonttitle=\bfseries\small, fontupper=\small,
  coltitle=blue!50!black, colbacktitle=blue!30!white,
  before skip=12pt, after skip=12pt
}

\title{Data Curation and Model Souping: Unlocking Qwen3-0.6B for MS MARCO Passage Ranking}

\author{%
  Shuguang Han \\
  Xianyu of Alibaba \\
  \texttt{shuguang.sh@alibaba-inc.com} \\
}

\begin{document}

\maketitle

\begin{abstract}
This report studies passage retrieval and re-ranking on the MS MARCO benchmark using lightweight Qwen3-0.6B models. Across three experimental settings --- BM25-then-reranking task, corpus-wide dense retrieval task, and dense retrieval then re-ranking task --- the results demonstrate that competitive ranking performance can be achieved without complex architectural modifications, custom loss functions, or cross-encoder distillation. Instead, substantial gains stem from data-centric strategies: corpus-targeted continued pre-training, rigorous data curation, hard-negative mining, training--inference distribution alignment, and weight merging via Model Soup. In particular, Model Soup delivers marked performance gains while preserving the inference efficiency of a single model. These findings highlight how data curation and weight merging can push sub-billion-parameter models to state-of-the-art performance. My models can be downloaded from Hugging Face: https://huggingface.co/shuguangh/models
\end{abstract}

\section{Introduction}

MS MARCO passage ranking is an information retrieval task built on top of the MS MARCO dataset \citep{bajaj2016msmarco}. This dataset contains 1 million real-world Bing search queries (each framed as a question) and 8.8 million candidate passages. For each query, human annotators have labeled zero or more relevant passages. Prior work has primarily focused on two benchmark tasks: passage re-ranking and passage full ranking.
\begin{itemize}[leftmargin=1.2em, itemsep=3pt, topsep=3pt]\small
  \item \textbf{Passage Re-ranking}: For each query, systems are provided with the top 1,000 candidate passages retrieved by BM25. The goal is to re-rank these candidates based on their relevance to the query---specifically, the likelihood of a passage answering the question.
  \item \textbf{Passage Full Ranking}: While re-ranking performance is inherently bounded by the recall of the top 1,000 BM25 results, the full ranking task requires retrieving and ranking relevant passages directly from the entire corpus of 8.8 million documents. In practice, this setup typically requires developing both a first-stage retrieval model and a second-stage re-ranking model.
\end{itemize}

For MS MARCO retrieval and re-ranking, BERT-family models have long served as the dominant paradigm. Most existing studies build upon BERT \citep{devlin2019bert,han2020tfranking,bendersky2020rrf102}, RoBERTa \citep{liu2019roberta}, ELECTRA \citep{clark2020electra}, or ERNIE \citep{qu2021rocketqa,ren2021rocketqav2}, with some---such as SimLM \citep{wang2023simlm} and CoT-MAE \citep{wu2023cotmae}---further exploring specialized pre-training objectives. More recently, with the rapid advancement of Large Language Models (LLMs), recent studies have begun examining the utility of state-of-the-art generative language models such as T5 and LLaMA for superior ranking performance \citep{pradeep2023rankzephyr,ma2024finetuningllama,zhang2025qwen3embedding}.

In this report, I follow this line of work by leveraging LLMs to improve ranking quality. Specifically, in collaboration with Claude Code (powered by Claude Opus 4.7), I explore effective strategies for utilizing the Qwen3-0.6B model \citep{zhang2025qwen3embedding} in both passage retrieval and re-ranking. I find that even without special pre-training, distillation from larger teacher models, loss function modifications, or architectural changes, rankers and retrievers built on Qwen3-0.6B can achieve state-of-the-art performance solely through deliberate training data curation and model merging. In particular, the Model Soup algorithm \citep{wortsman2022modelsoups}---which averages weights across multiple fine-tuned model checkpoints---proves to be an exceptionally effective strategy for boosting model capability.

During this research, I also noticed that performance metrics reported across existing studies are not directly comparable: while most prior works omit passage titles, a subset incorporates them. In my experiments, appending passage titles consistently yields an improvement of 2--3 percentage points (pp) in MRR@10 for retrieval and 1--2 pp for re-ranking. This discrepancy aligns with observations previously noted by \citet{lassance2023twomsmarco}. Consequently, to ensure fair and rigorous comparisons, all baseline metrics from prior literature in this report are systematically reproduced using their open-source checkpoints.

Another noteworthy aspect of this work is that throughout the entire project, I did not write a single line of code myself. In fact, given the sheer volume of code generated, comprehensively inspecting its correctness by hand was practically impossible. Claude Code provided actionable guidance during execution and demonstrated strong capabilities in diagnosing model performance issues. For instance, upon identifying the discrepancy between published results and my reproduced results, Claude Code deduced the inconsistent usage of title content and autonomously designed an experiment suite to validate the hypothesis. Nevertheless, the high-level intuition and direction behind each experiment still relied on my input, as Claude Code tended to stick to more generic solutions.

\section{Methodology}

In this technical report, I study three distinct experimental setups: (1) the BM25-then-Reranking pipeline, which re-ranks the top 1,000 candidate passages retrieved by BM25; (2) the Dense Retrieval task, which retrieves the most relevant passages directly from the entire corpus; and (3) the DenseRetrieval-then-Reranking pipeline, which re-ranks the top 1,000 passages retrieved by an initial dense retriever.

\subsection{Backbone Models}

All of my experiments are conducted by fine-tuning two Qwen3-0.6B based models \citep{zhang2025qwen3embedding}: Qwen3-0.6B-Embedding for retrieval and Qwen3-0.6B-Reranker for re-ranking. According to their technical report \citep{zhang2025qwen3embedding}, both models are developed via multi-stage training pipelines built on top of the Qwen3 foundation model, implementing dual-encoder and cross-encoder architectures, respectively.

More specifically, both models are initialized directly from the pre-trained Qwen3-0.6B language model, thereby inheriting strong multilingual, long-context capabilities. The Embedding model adopts a dual-encoder architecture optimized through contrastive pre-training, supervised fine-tuning on high-quality text pairs, and model merging to maximize generalization. It extracts semantic representations from the final end-of-sequence token and supports Matryoshka Representation Learning (MRL) for flexible vector dimensions. The Reranker model employs a cross-encoder that jointly processes query-document pairs via deep cross-attention layers. It is trained with supervised fine-tuning on labeled relevance pairs to predict query-passage relevance.

Due to GPU resource constraints, I restrict my focus in this report to sub-1B models, leaving larger-sized models such as Qwen3-Embedding-4B/8B and Qwen3-Reranker-4B/8B for future exploration. Nevertheless, I expect that ranking performance could be scaled further with larger parameter sizes, as observed in the original Qwen3 Embedding report \citep{zhang2025qwen3embedding}.

\subsection{Data Preparation}

\paragraph{Datasets.} I evaluate on the MS MARCO passage ranking benchmark \citep{bajaj2016msmarco}, which contains 501,233 queries in the training set and 6,800 queries in the development set, evaluated over a corpus of more than 8.8 million passages. In this dataset, each query is paired with one or more human-labeled relevant passages. The official benchmark provides the top 1,000 candidate passages per query retrieved by an un-tuned BM25 system (which can be substantially improved with hyperparameter tuning \citep{nogueira2019multistage}).

\paragraph{Training Data Construction.} For each of the three tasks mentioned above, I construct dedicated training datasets as follows:

The BM25-then-Reranking task: The dataset is constructed in two steps. First, I sample $N$ queries from the training partition. Next, for each sampled query, I take the top 1,000 passages retrieved by the official BM25 model, keep all labeled relevant passages as positive samples, and randomly sample $M$ irrelevant passages as negatives. Here, $N$ (query count) and $M$ (negative samples per query) are the primary parameters controlling the training scale. A representative data record is formatted as follows:

\begin{databox}
[system]
Judge whether the Document meets the requirements based on the Query
and the Instruct provided. Note that the answer can only be "yes"
or "no".

[user]
<Instruct>: Given a web search query, retrieve relevant passages
that answer the query
<Query>: what kind of organism is a black damsel
<Document>: Body of the organism is made up of a single cell. The
body of mutlicellular organism is made up of numerous cells. Body
organization is simple. Organization is complex ...

[assistant]

no
\end{databox}

The Retrieval task: The goal is to retrieve relevant passages directly from the full 8.8-million-passage collection. To maintain low inference latency, the dual-encoder architecture is preferred over the cross-encoder for corpus-wide similarity search. The data format for training and inference is illustrated below. To boost retrieval accuracy, my focus---consistent with prior studies \citep{formal2022splade,ni2022largedualencoders,wang2023simlm,santhanam2022colbertv2}---primarily centers on effective hard negative mining strategies. Specifically, I follow conventional practice by sampling hard negatives from candidates ranked between positions 20 and 200 by an existing retrieval model or cross-encoder reranker.

\begin{databox}
Query side:   Instruct: Given a web search query, retrieve relevant
              passages that answer the query
              Query: what kind of organism is a black damsel<|endoftext|>
Passage side: Body of the organism is made up of a single cell. The
              body of multicellular ...<|endoftext|>
\end{databox}

The DenseRetrieval-then-Reranking task: Similar to the BM25-then-Reranking setup, I first build an initial candidate pool using the trained dense retrieval model (retrieving the top 1,000 passages for all training queries). Then, I sample $N$ training queries, retaining all relevant passages as positives and sampling $M$ negative passages from the remaining non-relevant candidates. The resulting data schema is exactly the same as that used in the BM25-then-Reranking pipeline.

\subsection{Experiment Setup}

\paragraph{Implementation Details.} In this report, all of my experiments are trained and evaluated on an $8 \times 16$GB NVIDIA Tesla V100 GPU cluster. Due to hardware resource constraints: (1) for the re-ranking task, I set the maximum sequence length to 384 and the effective batch size to 256; (2) for the retrieval task, I cap the maximum query and document lengths at 64 and 144 tokens, respectively, with effective batch sizes varying across different retrieval models. Following standard language model fine-tuning practices, I adopt the AdamW optimizer with a default learning rate of $1\mathrm{e}{-5}$.

\paragraph{Evaluation Metrics.} Both the retrieval and re-ranking tasks are evaluated using three metrics: Recall@K ($K \in \{100, 200, 1000\}$), MRR@10, and nDCG@10. Formal definitions of these metrics can be found in \citet{han2020tfranking}. In general, Recall primarily assesses corpus-wide retrieval coverage, whereas MRR and nDCG quantify top-rank precision for the re-ranking tasks.

\section{Experimental Results}

\subsection{Reranking the official BM25 retrieval result (BM25-then-Reranking)}

In this section, my primary focus is to re-rank BM25-retrieved results using models based on Qwen3-0.6B-Reranker. I first evaluate the zero-shot ranking performance of the original Qwen3-0.6B-Reranker (without fine-tuning). For a fairer comparison, I include fine-tuned MonoBERT \citep{nogueira2019multistage} and fine-tuned RankT5-Large \citep{zhuang2023rankt5} as strong baselines. As shown in Table~\ref{tab:bm25-rerank-baselines}, the vanilla Qwen3-0.6B-Reranker achieves an MRR@10 of 0.3289, clearly demonstrating its capability to identify relevant passages beyond the initial BM25 ranking. Furthermore, even with a relatively small training dataset (a total of 100K queries, pairing each query with 1--2 relevant and 10 irrelevant passages), Supervised Fine-Tuning (SFT) yields a substantial performance boost, nearly matching the top-performing non-ensembled models on the leaderboard.

In the following sections, I detail further explorations aimed at improving model performance. As an additional reference, I also experiment with RankT5-3B, which achieves an 0.3997 in MRR@10 and an 0.4586 in nDCG@10. However, due to the GPU memory constraints, I could not conduct further experiments with the larger Qwen3-Reranker-4B/8B variants. Nevertheless, these results strongly suggest that ranking performance could be further scaled up alongside model parameter size.

\begin{table}[t]
  \caption{Performance comparison across baseline models (BM25, MonoBERT, TFR-BERT, RankT5-Large) and Qwen3-0.6B-Reranker based fine-tuned models in terms of MRR@10 and nDCG@10.}
  \label{tab:bm25-rerank-baselines}
  \centering
  \resizebox{\textwidth}{!}{%
  \begin{tabular}{lccc}
    \toprule
     & Model Size & MRR@10 & nDCG@10 \\
    \midrule
    BM25 & - & 0.1876 & 0.2340 \\
    MonoBERT-large \citep{nogueira2019multistage} & 0.34B & 0.3666 & 0.4262 \\
    TFR-BERT Single Run \citep{han2020tfranking} & 0.34B & 0.3782 & - \\
    RankT5-large \citep{zhuang2023rankt5} & 0.77B & 0.3851 & 0.4453 \\
    monoBERT + duoBERT (leaderboard best for non-ensemble) \citep{nogueira2019multistage} & 0.11B ($\times$2 models) & \textbf{0.3900} & - \\
    TFR-BERT Ensemble (Leaderboard best ensemble) \citep{han2020tfranking} & 0.34B ($\times$15 models) & \textbf{0.4046} & - \\
    \addlinespace
    Qwen3-0.6B-Reranker & 0.60B & 0.3289 & 0.3911 \\
    Qwen3-0.6B-Reranker-SFT (100K queries, w/o TCP) & 0.60B & 0.3817 & 0.4399 \\
    \bottomrule
  \end{tabular}%
  }
\end{table}

\paragraph{Target Corpus Pre-training (TCP).} \citet{nogueira2019multistage} developed a multi-stage document re-ranking pipeline where the initial stage involves continued pre-training of BERT on the MS MARCO corpus. Inspired by this idea, I also conduct Target Corpus Pre-training (TCP), continuing language model pre-training directly on the MS MARCO dataset. Specifically, I perform TCP on the Qwen3-0.6B-Reranker model using a standard causal language modeling (CLM) objective---namely, next-token prediction loss. The training corpus is the MS MARCO passage collection, pre-tokenized offline into fixed-length chunks of 512 tokens. Consistent with \citet{nogueira2019multistage}, I observe a clear incremental performance gain when adopting TCP, and this performance boost remains consistent across different training data scales. Throughout this technical report, this TCP-adapted model serves as the backbone for all downstream fine-tuned rerankers.

\begin{table}[t]
  \caption{Impact of Target Corpus Pre-training on fine-tuned model evaluated by MRR@10 and nDCG@10.}
  \label{tab:tcp}
  \centering
  \resizebox{\textwidth}{!}{%
  \begin{tabular}{lcccccccc}
    \toprule
     & SFT 10K & \makecell{SFT 10K \\ w/ TCP} & SFT 50K & \makecell{SFT 50K \\ w/ TCP} & SFT 100K & \makecell{SFT 100K \\ w/ TCP} & SFT 200K & \makecell{SFT 200K \\ w/ TCP} \\
    \midrule
    MRR@10 & 0.3618 & 0.3626 (+0.22\%) & 0.3749 & 0.3771 (+0.59\%) & 0.3817 & 0.3832 (+0.39\%) & 0.3835 & 0.3846 (+0.29\%) \\
    nDCG@10 & 0.4204 & 0.4211 (+0.17\%) & 0.4338 & 0.4352 (+0.32\%) & 0.4399 & 0.4418 (+0.43\%) & 0.4433 & 0.4441 (+0.18\%) \\
    \bottomrule
  \end{tabular}%
  }
\end{table}

\paragraph{Supervised Fine-tuning (SFT) with Varying Training Data Scales.} To understand how the volume of training data affects fine-tuning performance, I experiment with training datasets of varying sizes. These subsets are created by sampling queries and their corresponding passages from the full corpus: specifically, I sample $M$ queries, pairing each with $N$ relevant and $K$ irrelevant passages. In the experiments below, I evaluate $M \in \{100\mathrm{K}, 300\mathrm{K}, 500\mathrm{K}\}$, with $N{=}1$ and $K \in \{10, 14, 18\}$. Notably, 500K queries cover nearly the entire corpus. As shown in Table~\ref{tab:sft-scales} below, supervised fine-tuning significantly boosts the model's ranking ability, with performance exhibiting an upward trend as more training queries are introduced. In addition, I examine different numbers of negative samples ($K{=}10$, 14, and 18), revealing that $K{=}14$ yields the optimal performance, whereas a further increase in negative samples degrades the metrics. Notably, the MRR@10 of 0.3952 already outperforms the best non-ensemble baseline in Table~\ref{tab:bm25-rerank-baselines}.

\begin{table}[t]
  \caption{Performance of fine-tuned Qwen3-0.6B-Reranker under varying training data scales. Here, $M$ denotes the number of sampled queries, and $K$ denotes the number of sampled irrelevant passages per query.}
  \label{tab:sft-scales}
  \centering
  \resizebox{\textwidth}{!}{%
  \begin{tabular}{lcccccc}
    \toprule
     & \makecell{Base \\ (w/o SFT)} & \makecell{Model 1 \\ (M=100K, K=10)} & \makecell{Model 2 \\ (M=300K, K=10)} & \makecell{Model 3 \\ (M=500K, K=10)} & \makecell{\textbf{Model 4} \\ \textbf{(M=500K, K=14)}} & \makecell{Model 5 \\ (M=500K, K=18)} \\
    \midrule
    MRR@10 & 0.3289 & 0.3832 & 0.3859 & 0.3922 & \textbf{0.3952} & 0.3934 \\
    nDCG@10 & 0.3911 & 0.4418 & 0.4441 & 0.4509 & \textbf{0.4537} & 0.4534 \\
    \bottomrule
  \end{tabular}%
  }
\end{table}

\paragraph{Model Ensemble (ME).} Beyond individual fine-tuning, I employ model ensembling to boost ranking performance \citep{chen2025llmensemble}. Following the same sampling strategy (sampling $M$ queries, each with $N$ relevant and $K$ irrelevant passages), I generate six distinct datasets and train six independent models using identical configurations. I then experiment with two ensemble strategies: prediction-level ensemble \citep{chen2025llmensemble} via score fusion, and weight-space ensemble (the Model Soup algorithm \citep{wortsman2022modelsoups}), which averages model weights across multiple runs. The resulting soup model, \textbf{BR1}, surpasses the strongest single-model baseline by 2.74\%, and the prediction ensemble achieves performance on par with the TFR-BERT ensemble \citep{han2020tfranking} using just six runs. These results validate both the utility of Qwen3-0.6B-Reranker for fine-tuning and the effectiveness of my ensemble pipeline.

\begin{table}[t]
  \caption{Performance summary of different approaches. Improvements are calculated relative to the average baseline performance.}
  \label{tab:ensemble-rerank}
  \centering
  \resizebox{\textwidth}{!}{%
  \begin{tabular}{lccccc}
    \toprule
     & \makecell{Average performance \\ (of 6 models)} & \makecell{Model soup* \citep{wortsman2022modelsoups} \\ (on 3 models)} & \makecell{Result ensemble \\ (on 3 models)} & \makecell{\textbf{BR1}: Model soup* \citep{wortsman2022modelsoups} \\ (on 6 models)} & \makecell{Result ensemble \\ (on 6 models)} \\
    \midrule
    MRR@10 & 0.3952 & 0.3992 (+1.01\%) & 0.4019 (+1.70\%) & \textbf{0.4007 (+1.39\%)} & \textbf{0.4050 (+2.48\%)} \\
    nDCG@10 & 0.4537 & 0.4578 (+0.90\%) & 0.4600 (+1.39\%) & \textbf{0.4592 (+1.21\%)} & \textbf{0.4629 (+2.03\%)} \\
    \bottomrule
  \end{tabular}%
  }
  \par\vspace{4pt}
  \parbox{\textwidth}{\footnotesize Note: The model soup algorithm creates a single model with weights averaged from individual models; therefore, its inference resource consumption is equivalent to that of a single model. The \textbf{BR1} model can be accessed from Hugging Face: https://huggingface.co/shuguangh/bm25-rerank-model-soup.}
\end{table}

\begin{insightbox}{Takeaway}
Across the experiments above, two conclusions stand out: first, combining target corpus pre-training, data scaling, and model merging allows my single model to beat the previous best single-model baseline (0.4007 vs.\ 0.3900 MRR@10), even slightly surpassing the 3B RankT5 model (0.3997). Second, ensembling only a few models achieves performance on par with the best ensemble on the leaderboard (0.4050 vs.\ 0.4046 MRR@10).
\end{insightbox}

\subsection{Retrieving the whole data corpus (the Dense Retrieval task)}

For a given query, the retrieval task retrieves the most relevant passages from the entire corpus (around 8.8 million passages). To ensure low inference latency, industrial systems commonly adopt the dual-encoder architecture \citep{gao2022cocondenser,qu2021rocketqa} that encodes queries and documents separately and computes their dot-product similarity for online inference. In this subsection, I adopt the dual-encoder as the default architecture, and using Qwen3-0.6B-Embedding as the backbone model. To assess the capability of this backbone and evaluate my subsequent improvements, I benchmark against various representative baselines, including DeepCT \citep{dai2019deepct}, GTR-Large \citep{ni2022largedualencoders}, CoROM-Retriever \citep{corommodelscope}, coCondenser \citep{gao2022cocondenser}, SPLADE-coCondenser \citep{formal2022splade}, SimLM \citep{wang2023simlm}, CoT-MAE \citep{wu2023cotmae}, and ColBERTv2 \citep{santhanam2022colbertv2}.

As shown in Table~\ref{tab:retrieval-baselines}, even without supervised fine-tuning, the vanilla Qwen3-0.6B-Embedding model achieves a Recall@1000 of 0.9799 and an MRR@10 of 0.3165. Fine-tuning on the full training query set (pairing each query with 23 GTR-Large hard negatives) allows it to surpass GTR-Large across both MRR and nDCG metrics, demonstrating the clear efficacy of fine-tuning the Qwen3-0.6B-Embedding backbone. Among all baseline models, SimLM \citep{wang2023simlm} and ColBERTv2 \citep{santhanam2022colbertv2} yield the strongest performance. Specifically, SimLM introduces a replaced language modeling pre-training objective alongside a multi-stage pipeline featuring hard negative mining and cross-encoder distillation. Meanwhile, albeit at the expense of higher inference latency and storage footprint, ColBERTv2 adopts the late-interaction paradigm---independently encoding query and document tokens via BERT and aggregating relevance through MaxSim operators---which delivers the highest overall scores in Table~\ref{tab:retrieval-baselines}. Finally, while prior work \citep{ma2024finetuningllama} showed that the LLaMA-7B representation model achieves state-of-the-art performance on both Recall@1000 (0.9940) and MRR@10 (0.4120), I did not evaluate it in this technical report due to GPU memory constraints.

\begin{table}[t]
  \caption{Performance comparison across BM25, Qwen3-0.6B-Embedding, and representative dense retrieval baselines in terms of Recall@1000, MRR@10, and nDCG@10.}
  \label{tab:retrieval-baselines}
  \centering
  \resizebox{\textwidth}{!}{%
  \begin{tabular}{lcccc}
    \toprule
    \textbf{Retrieval} & \textbf{Model Size} & \textbf{Recall@1000} & \textbf{MRR@10} & \textbf{nDCG@10} \\
    \midrule
    BM25 - official \citep{bajaj2016msmarco} & - & 0.8140 & 0.1670 & - \\
    BM25 - optimized \citep{pyserini2021} & - & 0.8690 & 0.1876 & 0.2340 \\
    DeepCT \citep{dai2019deepct} & - & 0.9100 & 0.2430 & - \\
    GTR-Large \citep{ni2022largedualencoders} & 0.34B & \textbf{0.9873} & 0.3623 & 0.4268 \\
    CoROM-Retriever \citep{corommodelscope} & 0.11B & 0.9799 & 0.3682 & 0.4323 \\
    coCondenser \citep{gao2022cocondenser} & 0.11B & 0.9779 & 0.3705 & 0.4337 \\
    CoT-MAE \citep{wu2023cotmae} & 0.11B & 0.9805 & 0.3715 & 0.4350 \\
    SPLADE-CoCondenser \citep{formal2022splade} & 0.34B & 0.9825 & 0.3809 & 0.4471 \\
    SimLM \citep{wang2023simlm} & 0.34B & 0.9819 & 0.3947 & 0.4585 \\
    ColBERTv2 (late interaction) \citep{santhanam2022colbertv2} & 0.11B & 0.9738 & \textbf{0.3968} & \textbf{0.4629} \\
    \addlinespace
    Qwen3-0.6B-Embedding & 0.60B & 0.9799 & 0.3165 & 0.3796 \\
    Qwen3-0.6B-Embedding (w/ 23 GTR-Large hard negatives) & 0.60B & 0.9801 & 0.3789 & 0.4437 \\
    \bottomrule
  \end{tabular}%
  }
  \par\vspace{4pt}
  \parbox{\textwidth}{\footnotesize Note: To ensure fair and consistent evaluation (eliminating confounding factors from data preprocessing and metric computation pipelines), all baselines --- including GTR-Large, coCondenser, SPLADE-coCondenser, SimLM, CoT-MAE, and ColBERTv2 --- are evaluated locally under an identical protocol. Due to an input discrepancy (evaluations in \citep{wang2023simlm,wu2023cotmae} incorporate passage title content, whereas our setup does not), the reproduced MRR@10 is 0.3947 for SimLM (vs.\ 0.4110 reported in \citep{wang2023simlm}) and 0.3715 for CoT-MAE (vs.\ 0.3940 reported in \citep{wu2023cotmae}). Similarly, evaluating the official GTR-Large checkpoint from Hugging Face \citep{gfsgtrt5large} yields comparable nDCG@10 but a 1.7 pp drop in MRR@10 compared to the number reported in \citep{ni2022largedualencoders}. This impact of title discrepancy was also highlighted by \citet{lassance2023twomsmarco}.}
\end{table}

\paragraph{Hard Negative Mining.} According to Table~\ref{tab:retrieval-baselines}, the vanilla Qwen3-0.6B-Embedding model does not deliver performance on par with other dense retrieval models. A primary reason is that those high-performing models were all thoroughly fine-tuned on the MS MARCO dataset. Therefore, I conduct a series of fine-tuning experiments in this section. For the retrieval task, hard negative mining plays a pivotal role \citep{khattab2020colbert,formal2022splade,ni2022largedualencoders}. Here, I experiment with several strategies to mine high-quality hard negatives, drawing from BM25, GTR-Large \citep{ni2022largedualencoders}, SPLADE-coCondenser \citep{formal2022splade}, SimLM \citep{wang2023simlm}, ColBERTv2 \citep{santhanam2022colbertv2}, and my best cross-encoder reranker (model \textbf{FR1} from Section~\ref{sec:dr-rerank}).

More specifically, for each query, I use all relevant passages as positives, and randomly sample 23 non-relevant passages ranked between positions 10 and 200 (retrieved by a dense retriever or a cross-encoder reranker) as hard negatives. In-batch negatives (easy negatives) are shared across other queries within the same training batch. Training is carried out with 6 queries per batch per GPU across $8 \times$ NVIDIA V100 GPUs, resulting in an effective batch size of 1,152 passages per iteration. Under this setup, the ratio of easy negatives to hard negatives is approximately 49:1.

The retrieval results across different hard negative mining strategies are summarized in Table~\ref{tab:hard-negatives}. Compared to the vanilla Qwen3-0.6B-Embedding baseline, mining hard negatives from stronger retrieval models (e.g., GTR-Large, SPLADE-coCondenser, SimLM, and ColBERTv2) consistently boosts MRR@10 and nDCG@10, while Recall@1000 remains relatively stable. Beyond standard dense retrievers, I also evaluate hard negatives mined by the latter cross-encoder reranker \textbf{FR1}, which achieves the highest Recall@1000, MRR@10, and nDCG@10 scores across all mining strategies.

\begin{table}[t]
  \caption{Performance comparison of different hard negative mining strategies in terms of Recall@1000, MRR@10, and nDCG@10.}
  \label{tab:hard-negatives}
  \centering
  \resizebox{\textwidth}{!}{%
  \begin{tabular}{lccc}
    \toprule
    \textbf{Retrieval} & \textbf{Recall@1000} & \textbf{MRR@10} & \textbf{nDCG@10} \\
    \midrule
    Qwen3-0.6B-Embedding & 0.9799 & 0.3165 & 0.3796 \\
    Qwen3-0.6B-Embedding (w/ 23 GTR-Large hard negatives) & 0.9801 & 0.3789 & 0.4437 \\
    Qwen3-0.6B-Embedding (w/ 23 SPLADE-CoCondenser hard negatives) & 0.9763 & 0.3777 & 0.4416 \\
    Qwen3-0.6B-Embedding (w/ 23 SimLM hard negatives) & 0.9803 & 0.3783 & 0.4436 \\
    Qwen3-0.6B-Embedding (w/ 23 ColBERT V2 hard negatives) & 0.9790 & 0.3784 & 0.4431 \\
    Qwen3-0.6B-Embedding (w/ 23 cross-encoder hard negatives) & \textbf{0.9827} & \textbf{0.3853} & \textbf{0.4501} \\
    \bottomrule
  \end{tabular}%
  }
\end{table}

\paragraph{Model Ensemble (ME).} Similar to the approach taken for the BM25 re-ranking model \textbf{BR1}, I employ model ensembling to further improve dense retrieval performance. Following the same sampling strategy used for cross-encoder hard negative mining (i.e., retaining all relevant passages as positives and randomly sampling 23 hard negatives per query), I generate six distinct datasets and train six independent models with identical configurations. I then apply the Model Soup algorithm \citep{wortsman2022modelsoups} by averaging the weights across all six models into a single checkpoint, denoted as \textbf{DR1}.

Retrieval results are presented in Table~\ref{tab:ensemble-retrieval}. Compared to GTR-Large, the soup model \textbf{DR1} achieves significantly higher MRR@10, nDCG@1000, Recall@100, and Recall@200, while maintaining a comparable Recall@1000 with only a marginal difference. Compared to ColBERTv2, \textbf{DR1} attains superior Recall@100, Recall@200, and Recall@1000 at the expense of slightly lower MRR and nDCG. Notably, \textbf{DR1} stands as the top-performing dual-encoder retrieval model among all evaluated baselines. Additionally, I experimented with combining the model soup strategy with MaxSim token-level late interaction (as done in ColBERTv2); the resulting model, \textbf{DR2}, delivers the highest MRR, nDCG, and Recall scores overall. However, to preserve architectural simplicity and minimize inference latency, DR1 remains my default dense retrieval backbone for the full ranking pipeline.

\begin{table}[t]
  \caption{Performance comparison of different model ensembling strategies for dense retrieval across Recall@K, MRR@10, and nDCG@10.}
  \label{tab:ensemble-retrieval}
  \centering
  \resizebox{\textwidth}{!}{%
  \begin{tabular}{lccccc}
    \toprule
    \textbf{Retrieval} & \textbf{Recall@100} & \textbf{Recall@200} & \textbf{Recall@1000} & \textbf{MRR@10} & \textbf{nDCG@10} \\
    \midrule
    GTR-Large \citep{ni2022largedualencoders} & 0.9084 & 0.9494 & \textbf{0.9873} & 0.3623 & 0.4268 \\
    ColBERTv2 (late interaction) \citep{santhanam2022colbertv2} & 0.9135 & 0.9455 & 0.9738 & 0.3968 & 0.4629 \\
    Qwen3-0.6B-Embedding (w/ 23 cross-encoder hard negatives) & 0.9083 & 0.9446 & 0.9827 & 0.3853 & 0.4501 \\
    \textbf{DR1}: Qwen3-0.6B-Embedding model soup (6 models $\to$ 1 model) & 0.9172 & 0.9528 & 0.9861 & 0.3952 & 0.4614 \\
    \textbf{DR2}: Qwen3-0.6B-Embedding model soup + MaxSim (as in \citep{santhanam2022colbertv2}) & \textbf{0.9188} & \textbf{0.9521} & 0.9855 & \textbf{0.3968} & \textbf{0.4630} \\
    \bottomrule
  \end{tabular}%
  }
    \par\vspace{4pt}
  \parbox{\textwidth}{\footnotesize Note: The \textbf{DR1} model can be accessed from: https://huggingface.co/shuguangh/dense-retrieval-model-soup.}
\end{table}

\begin{insightbox}{Takeaway}
In this subsection, I have explored multiple model optimization techniques---primarily hard negative mining and model ensembling---to enhance dense retrieval performance. Through extensive experimentation, fine-tuned retrievers built upon the Qwen3-0.6B-Embedding model, even without specialized pre-training or architectural modifications, achieve near-SOTA performance across both Recall@1000 (0.9861 from \textbf{DR1} vs.\ 0.9873 from the similarly sized GTR-Large) and MRR@10 (0.3952 from \textbf{DR1} vs.\ 0.3968 from ColBERTv2). Moreover, among all evaluated dual-encoder retrievers (excluding ColBERTv2 and \textbf{DR2}, which adopt the late-interaction structure), the souped models \textbf{DR1} stand out as the top-performing systems. Finally, while I did not evaluate larger models such as GTR-XL, GTR-XXL, and RepLLaMA-7B due to GPU resource constraints, I believe these empirical findings and core conclusions remain robust.
\end{insightbox}

\subsection{Reranking the dense retrieval results (the DS-then-Reranking task)}
\label{sec:dr-rerank}

For most of the dense retrieval models discussed above, Recall@1000 already exceeds 0.9800. Therefore, in this section, I focus primarily on re-ranking the top 1,000 candidate passages retrieved by the dense retriever---that is, the full ranking task. To rigorously assess subsequent model improvements, I benchmark against several representative reranker baselines, including ME-BERT \citep{luan2021sparsedense}, RankT5-large \citep{zhuang2023rankt5}, CoROM \citep{corommodelscope}, CoT-MAE \citep{wu2023cotmae}, Ensemble of TFR-BERTs \citep{han2020tfranking}, RocketQA v2 \citep{ren2021rocketqav2}, and HLATR \citep{zhang2022hlatr}. For a fair comparison, the results of existing models in Table~\ref{tab:full-ranking} are obtained by re-evaluating their publicly released checkpoints under my unified evaluation protocol. Entries marked with question marks denote models that could not be fully reproduced, either due to the unavailability of public checkpoints or ambiguities in the original input data (as noted by \citet{lassance2023twomsmarco}, some studies utilize passage titles whereas others do not).

I first evaluate the best BM25-then-Reranking model, BR1, on the top 1,000 candidates retrieved by DR1; the corresponding results are presented in Table~\ref{tab:full-ranking}. Compared to the fully reproduced baselines such as RankT5-large, CoROM, monoT5-large, and CoT-MAE Reranker, BR1 achieves highly competitive performance, attaining an MRR@10 of 0.4168 and an nDCG@10 of 0.4840. Although models like RocketQA v2, HLATR, and the ensemble of fine-tuned ERNIE and RoBERTa report higher MRR@10 scores, their results cannot be faithfully reproduced because their checkpoints are unavailable and their input configurations remain uncertain.

\begin{table}[t]
  \caption{Performance comparison of full ranking models in terms of MRR@10 and nDCG@10. Entries marked with question marks indicate results that could not be fully reproduced, either due to the lack of public checkpoints or ambiguities in the evaluation data setup (e.g., whether passage titles were included \citep{lassance2023twomsmarco}).}
  \label{tab:full-ranking}
  \centering
  \resizebox{\textwidth}{!}{%
  \begin{tabular}{llccc}
    \toprule
    \textbf{Retrieval} & \textbf{Reranker} & \textbf{Reranker Model Size} & \textbf{MRR@10} & \textbf{nDCG@10} \\
    \midrule
    BM25 - official \citep{bajaj2016msmarco} & - & - & 0.1670 & - \\
    BM25 - optimized \citep{pyserini2021} & - & - & 0.1876 & 0.2340 \\
    ME-BERT \citep{luan2021sparsedense} & ME-BERT \citep{luan2021sparsedense} & 0.34B & 0.3950? & ? \\
    GTR-Large \citep{ni2022largedualencoders} & RankT5-large \citep{zhuang2023rankt5} & 0.77B & 0.4047 & 0.4739 \\
    coCondenser \citep{gao2022cocondenser} & CoROM \citep{corommodelscope} & 0.11B & 0.4136 & 0.4793 \\
    GTR-Large \citep{ni2022largedualencoders} & monoT5-large \citep{pradeep2021expandomonoduo} & 0.77B & 0.4156 & 0.4843 \\
    RocketQA V2 \citep{ren2021rocketqav2} & RocketQA V2 \citep{ren2021rocketqav2} & 0.12B & 0.4190? & ? \\
    CoT-MAE Retriever \citep{wu2023cotmae} & CoT-MAE Reranker \citep{wu2023cotmae} & 0.34B & 0.4200 & 0.4853 \\
    DeepCT \citep{dai2019deepct} & Ensemble of TFR-BERTs \citep{han2020tfranking} & 0.34B ($\times$15 models) & 0.4213 & - \\
    coCondenser \citep{gao2022cocondenser} & HLATR (three-stage reranking) \citep{zhang2022hlatr} & 0.34B & 0.4370? & ? \\
    Condenser \citep{gao2022cocondenser} & Ensemble of finetuned ERNIE and RoBERTa & 0.34B ($\times$? models) & 0.4430? & ? \\
    \addlinespace
    Best Qwen3 Retriever \textbf{DR1} & Best BM25-then-Reranking Model \textbf{BR1} & 0.60B & 0.4168 & 0.4840 \\
    \bottomrule
  \end{tabular}%
  }
  \par\vspace{4pt}
  \parbox{\textwidth}{\footnotesize Note: To ensure a fair and consistent evaluation, baseline models such as RankT5-large, CoROM, monoT5-large and CoT-MAE are evaluated locally under an identical protocol. Due to input discrepancies (e.g., the exclusion of passage titles in my setup) \citep{lassance2023twomsmarco}, the reproduced MRR@10 for CoT-MAE is 0.4200 (vs.\ 0.4388 reported in \citep{wu2023cotmae}), whereas the reproduced MRR@10 for CoROM matches that reported in \citep{corommodelscope}. For monoT5-large, the reproduced MRR@10 is 0.4156, while prior literature does not report results for monoT5-large paired with a GTR-Large retriever.}
\end{table}

\paragraph{Training-Inference Alignment.} The model BR1 was trained on 14 randomly sampled non-relevant passages from BM25 retrieval results, which does not closely reflect the candidate distribution produced by dense retrievers. This discrepancy can lead to a training-inference misalignment problem and degrade downstream ranking quality. To address this issue, I adopt a sampling procedure similar to that of BR1: I retain all labeled relevant passages and randomly sample 14 non-relevant candidates from the top 1,000 dense retrieval outputs. Using this newly constructed dataset, I then train a DenseRetrieval-then-Reranking model. Table~\ref{tab:dr-rerank} presents the corresponding performance. Clearly, a reranker trained directly on dense retrieval candidates better aligns with the dense retriever's output distribution, allowing the DR-then-Reranking model to outperform the BM25-then-Reranking counterpart (improving MRR@10 from 0.4087 to 0.4178).

\paragraph{Model Ensemble (ME).} As with BR1, I apply the Model Soup algorithm to further enhance full ranking performance. Specifically, under identical training configurations, I repeat the training run across six distinct random seeds to produce six independent models. Averaging their weights via Model Soup yields model \textbf{FR1}, which achieves substantial gains (boosting MRR@10 from 0.4168 to 0.4237). Furthermore, since DR1 already delivers strong first-stage retrieval quality, I construct \textbf{FR2} by linearly combining the retriever score from DR1 with the reranker score from FR1, which yields additional gains across ranking metrics (lifting MRR@10 from 0.4237 to 0.4269). Prior work on HLATR \citep{zhang2022hlatr} showed that incorporating a multi-stage reranking framework further enhances ranking quality. Following this intuition, I integrate an additional HLATR reranker on top of FR2, achieving an additional performance gain (increasing MRR@10 from 0.4269 to 0.4296). Ultimately, my final model outperforms all fully reproduced baselines in Table~\ref{tab:full-ranking}, nearly reaching an MRR@10 of 0.4300.

\begin{insightbox}{Takeaway}
In this subsection, I focused on the full ranking task. Through extensive experimentation, three clear conclusions emerge: first, aligning the training and inference data distributions is crucial for reranker training---shifting the training negative pool from BM25 candidates to dense retrieval results yields a significant performance boost; second, applying Model Soup across multiple repeated training runs remains a highly effective strategy for boosting model performance; and third, ranking quality can be further amplified by combining retriever and reranker scores alongside a listwise post-reranker like HLATR. Notably, models FR1 through FR3 even outperform much larger 3B models such as RankT5-3B (which reports an MRR@10 of 0.4208), underscoring the strong parameter efficiency of my approach.
\end{insightbox}

\begin{table}[t]
  \caption{A comparison of multiple fine-tuned full ranking models using MRR@10 and nDCG@10.}
  \label{tab:dr-rerank}
  \centering
  \resizebox{\textwidth}{!}{%
  \begin{tabular}{llcc}
    \toprule
    \textbf{Reranker} & \textbf{Model Config} & \textbf{MRR@10} & \textbf{nDCG@10} \\
    \midrule
    Single BM25-then-Reranking Model & Randomly sample 14 negatives from BM25 results & 0.4087 & 0.4756 \\
    Single DR-then-Reranking Model & Randomly sample 14 negatives from DR1 results & 0.4178 & 0.4857 \\
    \textbf{BR1}: BM25-then-Reranking Model Soup & Weighted soup of six models (single model) & 0.4168 & 0.4840 \\
    \textbf{FR1}: DR-then-Reranking Model Soup & Weighted soup of six models (single model) & 0.4237 & 0.4909 \\
    \textbf{FR2}: FR1 + Retriever Score & Retriever score + reranker score & \textbf{0.4269} & \textbf{0.4943} \\
    \textbf{FR3}: FR2 + HLATR (as done in \citep{zhang2022hlatr}) & \makecell[l]{Retriever score + reranker score \\ + HLATR} & \textbf{0.4296} & \textbf{0.4966} \\
    \bottomrule
  \end{tabular}%
  }
  \par\vspace{4pt}
  \parbox{\textwidth}{\footnotesize Note: The \textbf{FR1} model can be accessed from: https://huggingface.co/shuguangh/dense-rerank-model-soup.}
\end{table}

\section{Conclusions}

In this report, I investigated passage retrieval and re-ranking on the MS MARCO benchmark using Qwen3-0.6B-based models. Across three evaluation settings---BM25-then-reranking, corpus-wide dense retrieval, and dense-retrieval-then-reranking---the results demonstrate that competitive ranking performance can be achieved without complex architectural modifications, task-specific pre-training objectives, cross-encoder knowledge distillation, or customized loss functions. Instead, the primary performance gains stem from continued pre-training on the target corpus, rigorous training data curation, hard-negative mining, training-inference distribution alignment, and weight-space model merging via Model Soup \citep{wortsman2022modelsoups}. In particular, Model Soup proves to be an exceptionally effective and cost-efficient strategy for boosting ranking quality while keeping inference latency strictly identical to that of a single model. Although evaluating larger models (such as the 4B and 8B variants of Qwen3-Embedding and Qwen3-Reranker) is left for future exploration, the present findings strongly suggest that further gains can be achieved by scaling model parameters in tandem with the data-centric and merging strategies explored here.

Another important finding concerns benchmark evaluation comparability. Existing MS MARCO results are often not directly comparable because certain prior works incorporate passage titles while others omit them. In my experiments, prepending passage titles consistently improves retrieval MRR@10 by approximately 2--3 percentage points and re-ranking MRR@10 by roughly 1--2 percentage points. Therefore, to ensure a fair comparison, I reproduced all major baselines using their publicly available checkpoints under a unified evaluation protocol. This observation is consistent with prior analyses by \citet{lassance2023twomsmarco}, highlighting the pressing need for transparent reporting of input preprocessing pipelines in information retrieval benchmarking.

Finally, this project offers an interesting case study in AI-assisted research. Throughout this work, I collaborated with Claude Code, powered by Claude Opus 4.7. The assistant proved highly effective in engineering-intensive tasks, including environment setup, model implementation, code debugging, data processing, experiment orchestration, and result analysis. It was particularly instrumental in diagnosing discrepancies between published and reproduced metrics by helping identify the inconsistent use of passage titles and designing targeted validation tests. However, the model frequently defaulted to generic solutions and exhibited limited research taste. Many of the most critical breakthroughs---such as applying Model Soup, aligning the reranker training distribution with dense retrieval outputs, and prioritizing data curation over architectural modifications---originated from human intuition rather than the assistant's suggestions. This aligns with recent observations that AI agents can dramatically accelerate engineering execution but still struggle with high-level research ideation and creative direction \citep{kirgis2026aiagents}.

\end{document}